\documentclass[twocolumn,pre,nofootinbib]{revtex4}

\newcommand{\beq}{\begin{eqnarray}}
\newcommand{\eeq}{\end{eqnarray}}

\usepackage{graphicx,amsmath,amssymb,tabularx}

\usepackage{color} 

\begin{document}

\title{Spin glass behavior in a random Coulomb antiferromagnet}
\author{J. Rehn}
\affiliation{Max Planck Institute for the Physics of Complex Systems,
N\"othnitzer Strasse 38, 01187 Dresden, Germany}
\author{R. Moessner}
\affiliation{Max Planck Institute for the Physics of Complex Systems,
N\"othnitzer Strasse 38, 01187 Dresden, Germany}
\author{A.~P.~Young}
\affiliation{Department of Physics, University of California, Santa Cruz, California 95064, USA}


\begin{abstract}
We study spin glass behavior in a random Ising Coulomb antiferromagnet in two and
three dimensions using Monte Carlo simulations. In two dimensions, we find a
transition at zero temperature with critical exponents consistent with those
of the Edwards Anderson model, though with large uncertainties. In three
dimensions, evidence for a finite-temperature transition, as occurs in the
Edwards-Anderson model, is rather weak. This may indicate that the sizes are
too small to probe the asymptotic critical behavior, or possibly that the
universality class is different from that of the Edwards-Anderson model and
has a lower critical dimension equal to three. 
\end{abstract}

\maketitle

\section{Introduction}
\label{sec:intro}
Most studies of spin glasses use a model of the
Edwards-Anderson~\cite{edwards:75} type in which the interactions are
short-range and have random sign.  However, it is argued that spin glass
behavior is more general and that the necessary ingredients are simply
\textit{randomness} and \textit{frustration}. Indeed an antiferromagnet on a
random graph (which has all interactions negative) is
found~\cite{krzakala:08a} to have spin glass behavior, in which 
disorder and
frustration arise from large loops in the graph. In this paper we also
study spin glass behavior in a disordered model with only anti-ferromagnetic
interactions, but of the long-range Coulomb-type.

In addition to clarifying the
general conditions under which spin glass behavior can occur, further motivation for
our work comes from experiment. In certain highly frustrated random magnets,
Schiffer and Daruka~\cite{schiffer:97} showed that new magnetic degrees of
freedom emerge, so-called ``orphan'' spins, which could potentially
undergo a glassy transition. Subsequently, two of us and
collaborators~\cite{sen:12,rehn:15} showed that the orphan spins have Coulomb
interactions between them, due to entropic effects, and performed 
numerical simulations on the resulting model. A final motivation for our
study is that
Villain~\cite{villain:77b} showed that antiferromagnetic Coulomb
interactions arise between effective Ising spins, called ``chiralities'', in an
XY (i.e.~2-component) spin model with frustration and speculated that this
could lead to a glassy transition.

The main question we address in this work is whether the
spin glass transition in the random Coulomb
antiferromagnet is in the same universality class as the Edwards-Anderson (EA)
spin glass. Microscopically they are very different. The EA model is
short-range and its interactions have random sign while the Coulomb antiferromagnet
has all interactions negative and is long-range.  However,
Ref.~\cite{rehn:15} showed that there is a screening mechanism in the
random Coulomb antiferromagnet so we might expect that the interactions driving
a spin glass transition are short-range. In addition,
both the EA model and the random Coulomb antiferromagnet have
disorder and frustration, so one might imagine that 
the universal behavior at
a spin glass transition could be the same. We will try to see if this is the
case
by numerical simulations. Our conclusion is that the data is
\textit{consistent} with this hypothesis, but other scenarios can not be ruled
out because there are strong corrections to finite-size scaling for the
range of sizes that we can study.

Reference \cite{rehn:15} performed Monte Carlo simulations on a Heisenberg
(i.e.~three-component) version of the random Coulomb antiferromagnet because
the orphan spins emerging in experimental frustrated quantum
magnets~\cite{schiffer:97,sen:12} are of the Heisenberg type. However, in
order to try to answer questions about the spin glass
universality class we prefer to
study the Ising (i.e.~one-component) version of the model. One reason is that
the updating algorithm is simpler and more efficient than for the Heisenberg case.
More important is that even for the EA model, the nature of the spin glass
transition in, say, three dimensions has been harder to elucidate for the
Heisenberg case than for the Ising case. This is partly because the transition
temperature is much lower and partly because there seem to be larger
corrections to finite-size scaling as well as complications due to additional
(chiral) degrees of freedom, see for example Refs.~\cite{banos:12,viet:09a}.
By contrast, the transition in the three-dimensional Ising EA spin glass is
much better understood, see Refs.~\cite{hasenbusch:08b,baity-jesietal:13}. By
using Ising rather than Heisenberg spins, and by some refinements to the Monte
Carlo method, we are able to study significantly larger sizes than in
Ref.~\cite{rehn:15}.

The plan of this paper is as follows. In Sec.~\ref{sec:model} we describe the
model and the numerical method used to simulate it. In
Sec.~\ref{sec:quantities} we explain
the finite-size scaling method used to investigate the transition, while in
Sec.~\ref{sec:results} we describe the results and interpret them for the cases of
dimension $d$ equal to 2 and 3. Our conclusions are summarized in
Sec.~\ref{sec:conclusions}.

\section{The Model}
\label{sec:model}

We study $N$ Ising spins, $S_i = \pm 1$, randomly placed on a $d-$dimensional
hypercubic lattice of size $L$ for $d=2$ and $3$. The concentration of spins is
therefore $x = N / L^d$. 
The Hamiltonian is given by
\begin{equation}
\mathcal{H} = -\sum_{\langle i, j \rangle} J(\mathbf{r}_{ij}) S_i S_j \, ,
\label{Ham}
\end{equation}
where the interactions $J(\mathbf{r}_{ij})$ are given by the lattice Green function
\begin{equation}
J(\mathbf{r}_{ij}) = -{a_d \over L^d} \sum_{\mathbf{k}}
{\cos\left(\mathbf{k}\cdot\mathbf{r}_{ij}\right) \over d - \sum_{\ell = 1}^d \cos
k_\ell} \, .
\label{Jij}
\end{equation}
The factor $a_d$ is introduced so that the large-distance limit has the
Coulomb form
\begin{subequations}
\begin{align}
J(\mathbf{r}_{ij}) &=  \log (r_{ij} / \mathcal{L}), \qquad(d=2),\\
&= -1/r_{ij}, \qquad\qquad\ \ \ \ (d=3),
\label{large_r}
\end{align}
\label{coulomb}
\end{subequations}
where $\mathcal{L}$ is a constant which can be chosen to be larger than any
$r_{ij}$ so the interactions are all antiferromagnetic. In fact, since we impose
the ``charge neutrality'' condition,
\begin{equation}
\sum_{i=1}^N S_i = 0,
\label{neut}
\end{equation}
the Hamiltonian is actually independent of $\mathcal{L}$. The numerical values
of $a_d$ are
\begin{subequations}
\begin{align}
a_2 &= \pi, \\
a_3 &= 2 \pi.
\end{align}
\end{subequations} 
Note that, since the \textit{positions} of the spins are random, the
interactions will be different for different samples, but always
antiferromagnetic. To reduce error bars coming from sample-to-sample
fluctuations we need to average over many, typically several hundred, samples.

We simulate this model using the Metropolis Monte Carlo method, modified as
follows to incorporate the charge neutrality
condition in Eq.~\eqref{neut}. A site $i$ is
chosen, either sequentially or at random, and then one of the $z$ nearest
sites to this, $j$ say, is chosen at random. If the spins on $i$ and $j$
are antiparallel they
are both flipped with the usual Metropolis probability, and otherwise no change is
made. Repeating this procedure $N$ times corresponds to one Monte Carlo sweep. An
earlier version of the code took both spins to be random, so with high
probability they are far away, but this leads to an acceptance probability that
decreases rapidly with increasing system size.

We incorporate parallel
tempering (replica exchange)~\cite{hukushima:96,marinari:98b} to speed up
equilibration at low temperatures. In this approach, simulations are done at
several temperatures for the same set of interactions and
global moves are performed in which 
entire spin configurations at neighboring temperatures are exchanged, with a
probability satisfying the detailed balance condition. We determine the
temperatures empirically by requiring that the acceptance ratios for global
moves are reasonable, typically of order $0.3$. Appropriate
temperatures can be estimated with
sufficient accuracy from relatively short test runs.

\section{Quantities Calculated and Finite-Size Scaling}
\label{sec:quantities}

The main object of interest is the spin glass order parameter $q$ defined as
the instantaneous overlap between the 
spin configurations in two copies of the system with the same interactions,
\begin{equation}
q = {1 \over N} \sum_{i=1}^N S_i^{(1)} S_i^{(2)} \, ,
\end{equation}
where $``(1)"$ and $``(2)"$ refer to the copies. From measurements of $q$ we
determine the spin glass susceptibility
\begin{equation}
\chi_{SG} = N [\langle q^2 \rangle]_\text{av}
\end{equation}
and the Binder ratio
\begin{equation}
g = {1 \over 2}\, \left(3 - {[\langle q^4 \rangle]_\text{av}
\over [\langle q^2 \rangle]_\text{av}^2} \right)
\, ,
\end{equation}
in which angular brackets refer to a Monte Carlo average for
a single sample and square brackets $[\cdots]_\text{av}$ refer
to an average over samples. It is also useful to define a
wavevector-dependent spin glass susceptibility by
\begin{equation}
\chi_{SG}(\mathbf{k}) = {1 \over N}\sum_{i, j} \left[\left\langle S_i^{(1)} S_j^{(1)}
S_i^{(2)} S_j^{(2)} \right\rangle\right]_\text{av}
\cos\left(\mathbf{k}\cdot \mathbf{r}_{ij}\right) \, ,
\end{equation}
from which one can determine a correlation length $\xi_L$ according to
\begin{equation}
\xi_L = {1 \over 2 \sin(q_\text{min})} \, \left({\chi_{SG}(0) \over
\chi_{SG}(\textbf{q}_\text{min})} - 1\right)^{1/2} \, ,
\end{equation}
where $\textbf{q}_\text{min}$ is the smallest non-zero wavevector,
i.e.~$\textbf{q}_\text{min} = (2\pi/L)(1, 0)$ in $d=2$ and $\textbf{q}_\text{min} =
(2\pi/L)(1, 0, 0)$ in $d=3$.

To investigate whether or not there is a spin glass phase transition it is
essential to use finite-size scaling (FSS), see for example Refs.~\cite{privman:90,banos:12}.
If there is a transition at $T=
T_c$, in which the bulk (i.e.~infinite system-size) correlation length
diverges with an exponent $\nu$, i.e.
\begin{equation}
\xi_\infty \propto (T - T_c)^{-\nu}, 
\end{equation}
then the Binder ratio, being dimensionless, will have the FSS form
\begin{equation}
g = \widetilde{g}\left( L^{1/\nu} (T - T_c) \right)\, ,
\label{gscale}
\end{equation}
so curves of $g$ against $T$ for different sizes \textit{intersect} at $T_c$.

The spin glass susceptibility, however, is not dimensionless and for an
infinite system size diverges at $T_c$ with an exponent $\gamma$, i.e.
\begin{equation}
\chi_{SG} \propto (T - T_c)^{-\gamma}, \qquad (L \to \infty),
\end{equation}
so its FSS form is
\begin{equation}
\chi_{SG} = L^{2 - \eta} \, \widetilde{\chi}\left( L^{1/\nu} (T - T_c) \right)\, ,
\label{chisgscale}
\end{equation}
where $\eta$ is related to the other exponents by
\begin{equation}
\gamma = (2 -\eta) \nu \, .
\label{gamma}
\end{equation}

The correlation length $\xi_L$ divided by the system size is also
dimensionless and so has the FSS form
\begin{equation}
{\xi_L \over L} = \widetilde{X}\left( L^{1/\nu} (T - T_c) \right)\, .
\label{xiscale}
\end{equation}

\section{Results}
\label{sec:results}

\subsection{Screening}
\label{sec:screen}

\begin{figure}[tb!]
\begin{center}
\includegraphics[width=\columnwidth]{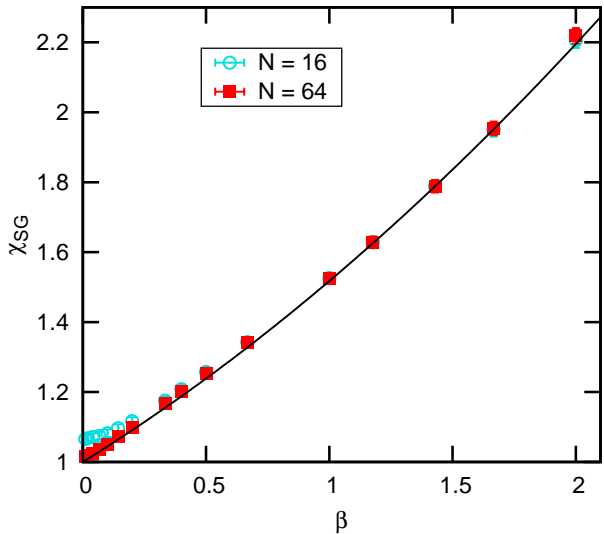}
\caption{
Data for the spin glass susceptibility at small $\beta \, (\equiv 1/T)$ in two
dimensions for a concentration $x = 1/16$ for $N = 16$ and $64$. The curve is
a fit to the data for $N = 64$.
The behavior is clearly linear, not
quadratic, at
small $\beta$, which is a result of screening. The intercept at $\beta = 0$
differs measurably from one for very small sizes only because of the charge
neutrality constraint, Eq.~\eqref{neut}.
\label{fig:screen}
}
\end{center}
\end{figure}

The spin glass susceptibility can be expressed as
\begin{align}
\chi_{SG} &= {1 \over N} \sum_{i,j} [C_{ij}^2]_\text{av} \nonumber \\
&= 1 + {1 \over N} \sum_{i \ne j} [C_{ij}^2]_\text{av}
\label{chi_Cij}
\end{align}
where $C_{ij} = \langle S_i S_j \rangle$ and the terms with $i = j$ give unity,
which is the result for $T = \infty$.  Naively, one can obtain the second term
in Eq.~\eqref{chi_Cij} at high temperature by expanding the Boltzmann factors
in powers of $\beta \, (\equiv 1/T)$, 
with the result
\begin{equation}
C_{ij} = \beta J_{ij} + O(\beta^2) \qquad (i \ne j) ,
\label{Cij}
\end{equation}
but then the sum in the second term in Eq.~\eqref{chi_Cij} diverges for the
Coulomb potential interaction in Eq.~\eqref{coulomb}. Clearly a resummation of
terms is needed to get a finite result. In fact,
Ref.~\cite{rehn:15} showed that the interactions are screened up to a length
scale $\lambda$ where
\begin{equation}
\lambda \propto \sqrt{T}.
\label{lambda}
\end{equation}
Within certain approximations, the final result of Ref.~\cite{rehn:15} in $d=2$, is

\begin{equation}
C_{ij} \propto \beta K_0(r_{ij} / \lambda) ,
\label{K0}
\end{equation}
where $K_0(x)$ is a modified Bessel functions which decays exponentially to
zero at large $x$. This is to be compared with 
the naive result in Eq.~\eqref{Cij} that
$C_{ij} = \beta J(\mathbf{r}_{ij}) \propto \beta
\log(r_{ij})$. Inserting Eq.~\eqref{K0} into Eq.~\eqref{chi_Cij} one has, at
high-$T$ and in $d = 2$,
\begin{align}
\chi_{SG} &= 1 + \text{const.}\, \lambda^2 \beta^2 \\
 &= 1 + \text{const.}' \, \beta \, .
\end{align}
Hence, because of screening, the leading correction to the
infinite-temperature result is of order $\beta$, rather than $\beta^2$ which
would be the case for short-range interactions. This result is clearly shown
by the numerics in Fig.~\ref{fig:screen}.

Because the interactions are screened, we might expect the universality class
of the spin glass transition in the disordered Coulomb antiferromagnet to be
the same as that of the \textit{short-range} EA spin glass. Even if this is
the case, the fact that the screening length is temperature dependent will
give rise to additional, and possibly 
large, corrections to FSS which could complicate the analysis. 

\subsection{Equilibration}
\label{sec:equil}
\begin{figure}[tb!]
\begin{center}
\includegraphics[width=\columnwidth]{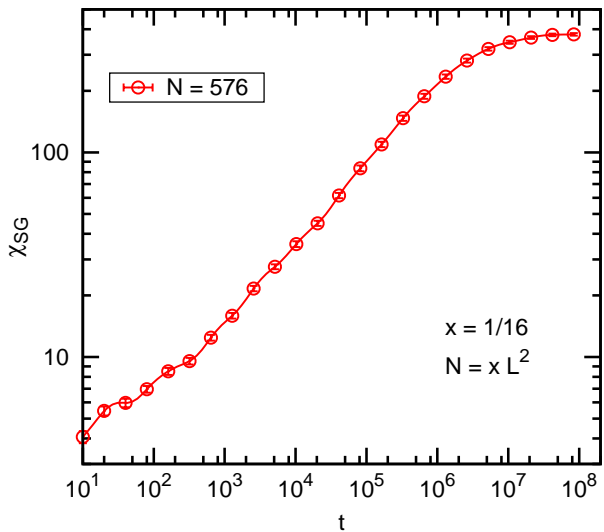}
\caption{
Equilibration plot for $N = 576, d=2$ with $x = 1/ 16$ at the lowest
temperature of $T = 0.032$. The spin glass susceptibility $\chi_{SG}$ is
plotted against Monte Carlo sweeps $t$.
For each point, averaging is carried out over the
last half of the sweeps. 
\label{fig:equil}
}
\end{center}
\end{figure}

To test for equilibration we obtain data for runs of different length in which
the number of sweeps doubles for each run, and for all runs we average over the
last half of the sweeps. It is easy to see that this can actually be done in a single
run by using \textit{all} the data. We require that the data is
independent of run time within small error bars for the last two data
points. Figure \ref{fig:equil} shows an example for $N = 576$ for $d = 2$ at
the lowest temperature $T = 0.032$.

\subsection{Two-dimensions}
\label{sec:2d}

\begin{table}[htb]
\caption{
Simulation parameters for $d=2$. The concentration is $x=1/16$ where $N = x
L^2$. For each value of 
size $N$, $N_{\mathrm{samp}}$ samples were run for $N_{\mathrm{sweep}}$
sweeps with
averaging performed over the last half.
Parallel tempering Monte Carlo was performed with $N_T$ temperatures distributed
between $T_{\mathrm{min}}$ and $T_{\mathrm{max}}$.
}
\begin{ruledtabular}
\begin{tabular}{rrrrrrr}
  $N$ & $L$ & $N_{\mathrm{sweep}}$ & $T_{\mathrm{min}}$ &
  $T_{\mathrm{max}}$ & $N_T$ & $N_{\mathrm{samp}}$ \\
  \hline
   64 & 32 & 2097152 & 0.025 & 1 & 15 & 256 \\
  144 & 48 & 2097152 & 0.025 & 0.500 & 15 & 1024 \\
  256 & 64 & 8388608 & 0.025 & 0.500 & 15 & 767 \\
  400 & 80 & 16777216 & 0.025 & 0.25 & 15 & 256 \\
  576 & 96 & 83886080 & 0.032 & 0.500 & 16 & 467 \\
  784 & 112 & 167772160  & 0.05 & 0.25 & 11 & 96 \\
\end{tabular}
\end{ruledtabular}
\label{tab:2d}
\end{table}

Now we discuss our results for two-dimensions for which we use a fixed
concentration $x = 1/16$. The parameters of the
simulations are shown Table~\ref{tab:2d}. 

For the EA model it is well established that the spin glass transition only
occurs at $T = 0$ where the correlation length diverges with an exponent $\nu
\simeq 3.4$~\cite{bray:84,hartmann:01a,fernandez:16} and the exponent
$\eta$, which is related to the
divergence of the spin glass susceptibility according to
Eqs.~\eqref{chisgscale}
and \eqref{gamma}, is
$\eta = 0$.

\begin{figure}[tb!]
\begin{center}
\includegraphics[width=\columnwidth]{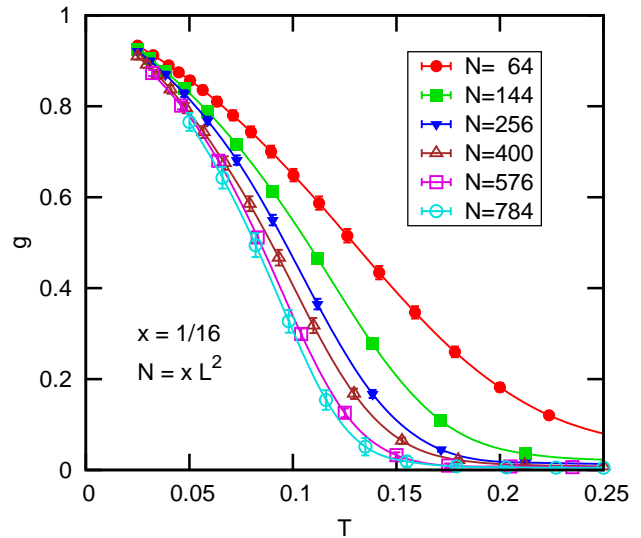}
\caption{
Data for the Binder ratio in $d = 2$.
\label{fig:g_2d}
}
\end{center}
\end{figure}

\begin{figure}[tb!]
\begin{center}
\includegraphics[width=\columnwidth]{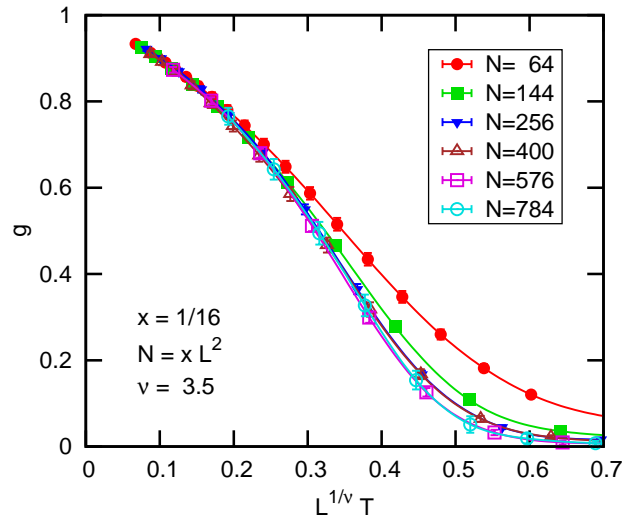}
\caption{
Scaling plot for Binder ratio in $d = 2$ according to Eq.~\eqref{gscale} with $T_c =
0$. We are not able to get \textit{all} the data to collapse for any value of
the correlation length exponent $\nu$. However the data for the largest sizes
collapses with $\nu \simeq 3.5$ (shown), consistent with the value in the EA
model. However, there are big uncertainties in our estimate (see text).
\label{fig:g_2d_scale}
}
\end{center}
\end{figure}

Our data for the Binder ratio $g$ is shown in Fig.~\ref{fig:g_2d}. There is no
sign of any intersections and hence no indication of a finite temperature
transition. This is consistent with the EA model in $d = 2$. Figure
\ref{fig:g_2d_scale} shows a scaling plot according to Eq.~\eqref{gscale} assuming
$T_c = 0$. If there is a zero temperature transition the data should collapse,
at least for large enough sizes and low enough temperatures.  We were not able
to collapse the data for 
\textit{all} sizes with any choice of the correlation length
exponent $\nu$ but the data for the largest sizes collapses fairly well with
$\nu\simeq 3.5$ (as shown) consistent with results for the EA model. However,
there is a large uncertainty in this estimate; we find that any value for
$\nu$ in the range $2.7$ to $4.5$ gives a plausible fit for large sizes. 

\begin{figure}[tb!]
\begin{center}
\includegraphics[width=\columnwidth]{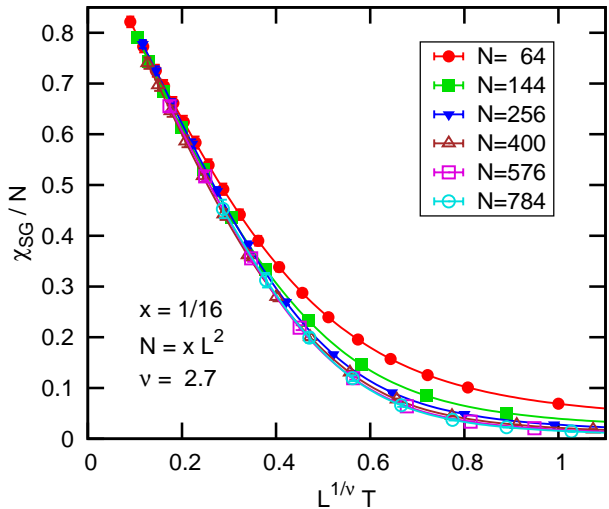}
\caption{
The spin glass susceptibility in $d = 2$ scaled according to
Eq.~\eqref{chisgscale} with $T_c = 0$ and $\eta = 0$. No value of $\nu$
succeeds in scaling all the data though the largest sizes scale reasonably
well with $\nu \simeq 2.7$ (shown), not very different from the value for the EA model
of $3.4$. However, there are big uncertainties in our estimate (see text).
\label{fig:chisg_2d}
}
\end{center}
\end{figure}

In Fig.~\ref{fig:chisg_2d} we show scaled 
data for $\chi_{SG}$ according to
Eq.~\eqref{chisgscale} assuming $T_c = 0$ and $\eta = 0$ (the latter
corresponding to a non-degenerate ground state which seems reasonable). 
As with the data for $g$ in
Fig.~\ref{fig:g_2d_scale}, we can not scale all the data for any value of
$\nu$.  However, the data for larger sizes scales fairly well with a value of
$2.7$ (shown) not very different from the value for the EA model of 3.4.
However, there are big uncertainties in our estimate; any value between $2.0$
and $4.0$ gives a plausible collapse for large sizes.

\begin{figure}[tb!]
\begin{center}
\includegraphics[width=\columnwidth]{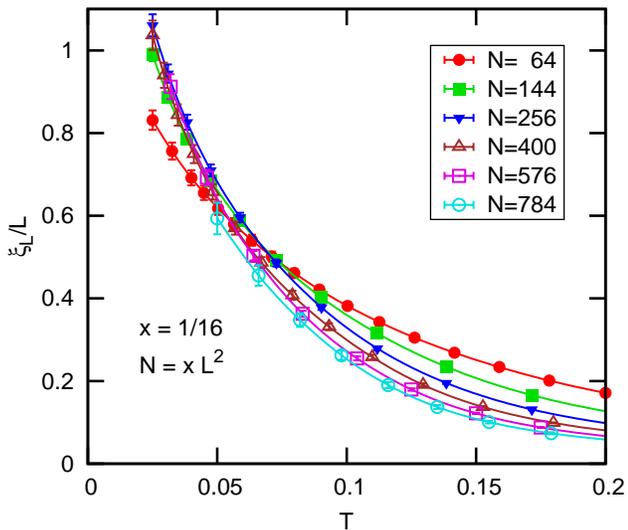}
\caption{
Correlation length divided by system size in $d = 2$.
\label{fig:xi_2d}
}
\end{center}
\end{figure}

\begin{figure}[tb!]
\begin{center}
\includegraphics[width=\columnwidth]{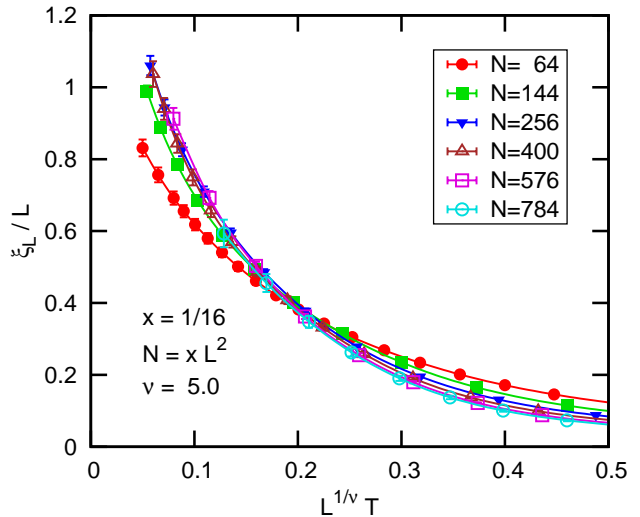}
\caption{
Scaling plot of the correlation length divided by system size in $d = 2$
assuming $\nu = 5.0$. 
\label{fig:xi_2d_scale}
}
\end{center}
\end{figure}

Our attempts to scale the data for $g$ and $\chi_{SG}$ indicate the presence
of substantial corrections to FSS for the range of sizes that we can study.
This problem is even more severe for the data for $\xi_L/L$. Being
dimensionless, the data for this quantity should
intersect if there is a transition, see Eq.~\eqref{xiscale}. As shown
in Fig.~\ref{fig:xi_2d} there \textit{is} an intersection involving the
smallest size studied, $N = 64$, but not for larger sizes. Rather the data for
larger sizes seems to merge at low-$T$.
If we try to scale the data for $\xi_L/L$ we need a
large value of $\nu$ to collapse the data for large sizes. Figure
\ref{fig:xi_2d_scale} shows the result with $\nu = 5.0$. 
These results indicate that the data for
$\xi_L/L$ is not at large enough sizes to give a clear prediction for the
nature of the transition. 

\subsection{Three-dimensions}
\label{sec:3d}

\begin{table}[htb]
\caption{
Simulation parameters for $d=3$. The concentration is $x=1/64$ where $N = x
L^3$. For each value of 
size $N$, $N_{\mathrm{samp}}$ samples were run for $N_{\mathrm{sweep}}$
sweeps with
averaging performed over the last half.
Parallel tempering Monte Carlo was performed with $N_T$ temperatures distributed
between $T_{\mathrm{min}}$ and $T_{\mathrm{max}}$.
}
\begin{ruledtabular}
\begin{tabular}{rrrrrrr}
  $N$ & $L$ & $N_{\mathrm{sweep}}$ & $T_{\mathrm{min}}$ &
  $T_{\mathrm{max}}$ & $N_T$ & $N_{\mathrm{samp}}$ \\
  \hline
   64 & 16 & 20480    & 0.0100 & 0.2000 &  9 & 1000 \\
  216 & 24 & 1310720  & 0.0100 & 0.2000 & 12 & 400 \\
  512 & 32 & 41943040 & 0.0200 & 0.2150 & 15 & 360 \\
 1000 & 40 & 83886080 & 0.0215 & 0.1305 & 15 & 568 \\
\end{tabular}
\end{ruledtabular}
\label{tab:3d}
\end{table}

Next we discuss our results for three-dimensions for which we use a fixed
concentration $x = 1/64$. The parameters of the
simulations are shown Table~\ref{tab:3d}. For the EA model in $d=3$ it is
firmly established that there is a spin glass transition at non-zero
temperature~\cite{hasenbusch:08b,baity-jesietal:13}.

\begin{figure}[tb!]
\begin{center}
\includegraphics[width=\columnwidth]{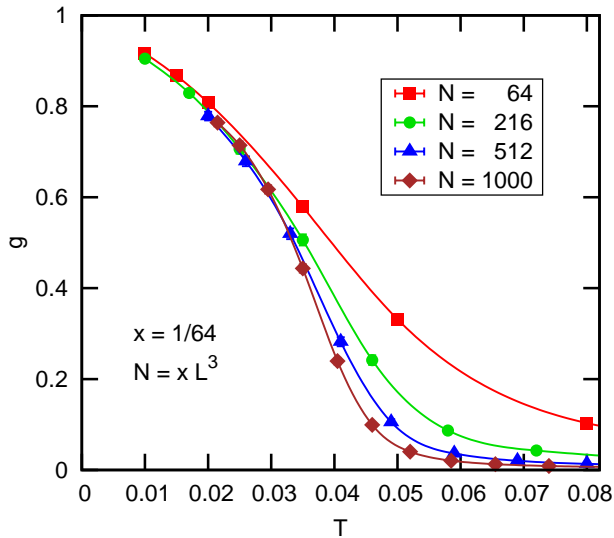}
\caption{
Binder ratio for $d = 3$.
\label{fig:g_3d}
}
\end{center}
\end{figure}

\begin{figure}[tb!]
\begin{center}
\includegraphics[width=\columnwidth]{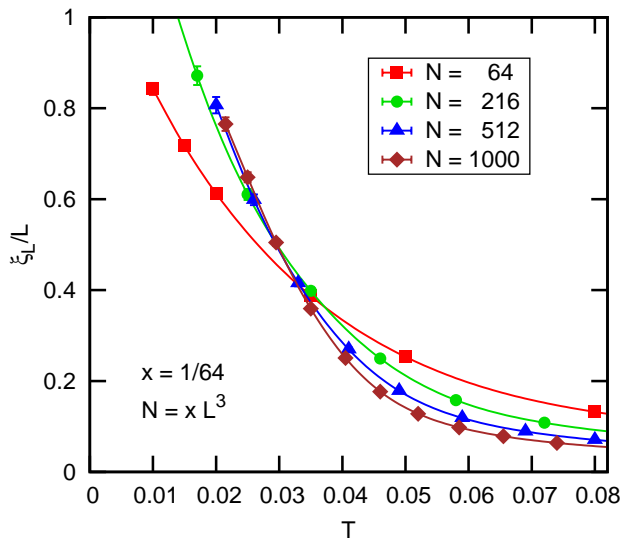}
\caption{
Correlation length divided by system size for $d = 3$.
\label{fig:xi_3d}
}
\end{center}
\end{figure}

Data for the dimensionless quantities $g$ and $\xi_L/L$ are shown in
Figs.~\ref{fig:g_3d} and \ref{fig:xi_3d} respectively.  The results for $g$
seem to merge at low-$T$ but do not obviously cross. It should be mentioned
that even for the EA model the splaying out of the data 
for $g$ below $T_c$ is only a small
effect, see for example Ref.~\cite{kawashima:96}, but is, nonetheless, observable
with good data on large sizes. 

For small sizes the data for $\xi_L/L$ shows a large splaying out, but the data
for large sizes seems only to merge. Splaying out for the smallest size was
also observed in $d=2$, see Fig.~\ref{fig:xi_2d}, and was interpreted as a FSS
correction since it disappears for larger sizes. The same is presumably true
here; we should give most weight to the data for larger sizes. But the larger
size data in Fig.~\ref{fig:xi_3d}
looks very marginal, possibly suggesting that $d = 3$ is the lower
critical dimension, $d_l$. This is different from the EA model where $d_l$ is
approximately, or possibly exactly~\cite{boettcher:05}, equal to $2.5$.

\section{Conclusions}
\label{sec:conclusions}
We have studied spin glass behavior in the random Coulomb Ising
antiferromagnet in two and three dimensions by Monte Carlo simulations, with
results analyzed by FSS. Since the interactions are screened, and so are
effectively short-ranged, a natural hypothesis is that the critical behavior
is the same as that of the short-range Ising EA model.
For the latter, a transition occurs at $T=0$ in $d=2$ but at a
non-zero temperature in $d = 3$. Our results indicate a zero
temperature transition in $d=2$, with a correlation length exponent compatible
with that found for the EA model, though with big error bars. However, in
$d=3$, we do not find unambiguous evidence for a non-zero temperature
transition temperature.
Rather the data for larger sizes seems to be ``marginal''. This could
indicate that the system sizes are simply not large enough to see the
asymptotic critical behavior, or it could be that our model is \textit{not} in
the same universality class as that of the short-range Ising EA model, but
rather has a different lower critical dimension, $d_l=3$ rather than $d_l=2.5$ for
the EA model. If this is the case, the nature of the physics causing the
difference in universal behavior is unclear to us.

There are clearly large corrections to FSS for this problem, the data for the
correlation length in Figs.~\ref{fig:xi_2d} and \ref{fig:xi_3d} providing
striking examples. As well as corrections to scaling that occur for the
short-range EA model, here we have an additional contribution because the
screening length is
\textit{temperature-dependent}. In fact, according to
Ref.~\cite{rehn:15} the screening length is singular for $T \to 0$, see
Eq.~\eqref{lambda}. If this result holds down to $T=0$ it could possibly
\textit{change} the critical behavior in $d=2$, where the transition is also
at zero temperature, rather than simply giving a \textit{correction} to scaling.
In $d=3$, the transition is at finite-$T$ for the EA model, so we expect only
a correction to scaling from the $T$-dependence of $\lambda$.

We close on a historical note. The question of whether or not there is a
finite temperature transition in the $d=3$ Ising EA spin glass was
controversial for many years. It was only later, when better FSS methods
were developed and
computers became more powerful, that the question was definitely answered in
the affirmative. Perhaps, therefore, it is not surprising that this early effort
on a Coulomb spin glass does not leave to a definite conclusion, given the
extra difficulties of long-range interactions and larger corrections to
scaling.

\begin{acknowledgments}
The work of APY is supported in part by the National
Science Foundation under Grant
No.~DMR-1207036 and by
a Gutzwiller Fellowship at the Max Planck Institute
for the Physics of Complex Systems (MPIPKS) Dresden. APY also thanks
Roderich Moessner for his kind hospitality while visiting the MPIPKS.
The work in Dresden was supported by DFG under grant SFB 1143.
RM and JR are grateful to Alex Andreanov, Kedar Damle, Anto Scardicchio and 
Arnab Sen for collaboration on related work. 
\end{acknowledgments}

\bibliography{refs}

\end{document}